\documentclass[aps,prb,twocolumn,superscriptaddress,floatfix]{revtex4}

\usepackage{amsmath,amssymb,bm}
\usepackage{graphicx}

\begin{document}

\title{Controlling Half-metallicity of Graphene Nanoribbons by Using a Ferroelectric Polymer}
\author{Yea-Lee \surname{Lee}}
\author{Seungchul \surname{Kim}}
\author{Changwon \surname{Park}}
\author{Jisoon \surname{Ihm}}
\email[Email:\ ]{jihm@snu.ac.kr}
\affiliation{Department of Physics and Astronomy, 
	        Center for Theoretical Physics,
             Seoul National University, 
             Seoul 151-747, Korea}
\author{Young-Woo \surname{Son}}
\email[Email:\ ]{hand@kias.re.kr}
\affiliation{Korea Institute for Advanced Study,
             Seoul 130-722, Korea}
\date{\today }
\begin{abstract}
On the basis of first-principles computational approaches,
we present a new method to drive zigzag graphene nanoribons (ZGNRs)
into the half-metallic state using a ferroelectric material, poly(vinylidene fluoride) (PVDF).
Owing to strong dipole moments of PVDFs, the ground state of the ZGNR becomes half-metallic
when a critical coverage of PVDFs is achieved on the ZGNR.
Since ferroelectric polymers are physisorbed, the direction of the dipole field
in PVDFs can be rotated by relatively small external electric fields, and
the switching between half-metallic and insulating states may be achieved.
Our results suggest that, without excessively large external gate electric fields,
half-metallic states of ZGNRs are realizable through the deposition of ferroelectric polymers
and their electronic and magnetic properties are controllable \textit{via} noninvasive mutual interactions.
\end{abstract}

\maketitle

Since graphene, a single layer of graphite, was shown to be exfoliated
on the silicon oxide surface~\cite{Novoselov},
there has been many studies on its various interesting physical, chemical,
and mechanical properties.
Specially, its unique electronic characteristics have spurred researchers
to envisage nanoelectronic circuits composed
of carbon atoms only~\cite{Berger,Geim}.
To achieve nanoelectronic devices based on graphene,
it should be cut into small pieces which inevitably have edges at the boundaries.
Together with the finite size effects~\cite{Wakabayashi,Nakada},
nanoscale graphene fragments with these edges exhibit many different properties
compared with those of two-dimensional graphene.
Thus, it is important to understand the physics of
the one-dimensional graphene nanostructure,
that is, graphene nanoribbon (GNR)~\cite{Wakabayashi,Nakada}.
Among various properties of the ribbon studied~\cite{Son1,Han,Chen,Li1,Wang,Son2,Lee,Jung},
the magnetism of the zigzag graphene nanoribbon (ZGNR) is especially
notable~\cite{Son2,Lee,Jung}.
The magnetism arises at the zigzag-shaped edges
because of the localized $\pi$-orbitals of carbon atoms
at the edges~\cite{Wakabayashi,Son2,Lee,Jung}.
Considering a quite long spin coherence length in graphene~\cite{Tombros},
the magnetism at the edges may be useful in future spintronics applications~\cite{Wolf}.

One of the most interesting properties regarding the application
of the magnetism in ZGNRs~\cite{Hod1,Hod2,Hod3,Cervantes,Li2,Kan,Zheng1}
is the electric-field-induced half-metallicity~\cite{Son2}.
Under the transverse electric field, ZGNRs show the half-metallicity~\cite{Son2}
which originates from a unique interplay between the interedge
antiferromagnetic ordering and the relative potential shift between two edges.
Despite its novel feature, the required transverse electric field
for the half-metallicity is too strong to be obtained easily in experiments~\cite{Son2}.
Many researchers have suggested diverse methods, such as introducing quantum dots
with the zigzag edges~\cite{Hod1,Hod2},
functionalizing edges of ZGNRs~\cite{Hod3,Cervantes,Li2},
and boron-nitrogen (B-N) substituting for designated carbon atoms~\cite{Kan,Zheng1}.
Some of these methods are successful in showing half-metallicity
without external electric fields, though impractical because
fine control of positions of functional groups and B-N substitution is inevitable.
Moreover, the half-metallic states are hardly controllable with those proposed methods.

In this paper, we propose a method to drive the ZGNR into a half-metallic state
by depositing a ferroelectric material, poly(vinylidene fluoride)
(PVDF)~\cite{Bune} on the ZGNR.
The PVDF, one of the well-known ferroelectric materials, shows strong ferroelectricity
and high crystallinity~\cite{Malin,Naber,Hu,Lovinger}.

The atomic model for the PVDF is in Figure 1a.
In general, two surfaces of the ferroelectric thin film, cut across the dipole moment aligned
to a certain direction, cause a potential difference due to the dipole field.
Likewise, several aligned PVDFs form the electric field which is induced
by the dipole moments, and even a few layers show a substantial ferroelectricity.
Here, we show that PVDFs can weakly bind to the surface of the ZGNR, with dipole
direction parallel to the surface, thereby achieving the half-metallicity
without any electric field. Because of their weak binding nature, the
direction of induced electric fields is controllable, and the switching
between half-metallic states and antiferromagnetic insulating states are made possible
by external gate fields.
We also note other experiment studies exploiting the interplay beetween ferroelectricity of the PVDF
and graphene~\cite{Zheng2, Ansari}.

\begin{figure}[t]
\centering
\includegraphics[width=1.00\columnwidth]{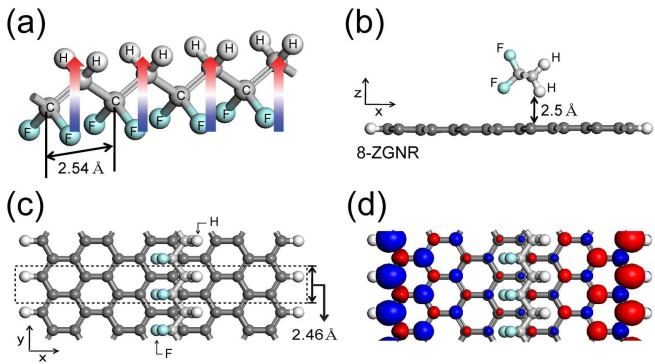}
\caption{
(a) Ball-and-stick model for Poly (vinylidene fluoride) (PVDF).
The arrows indicate the direction of the dipole moment (from the
fluorine side to the hydrogen side).
(b) Stable configuration of a PVDF chain deposited on the 8-ZGNR
whose edges are passivated by hydrogen atoms.
The PVDF chain and the ZGNR are set parallel along the $y$-axis.
The PVDF lies flat on the ZGNR, the height from the ZGNR to the lowest atom
of the PVDF is 2.5 \AA, and the PVDF sits on the center of the ZGNR.
(c) Top view of the same configuration as in (b).
(d) The isosurface of the charge density difference between the up-spin
and the down-spin ($\rho_{\uparrow}(r) - \rho_{\downarrow}(r)$)
when the PVDF is deposited on the 8-ZGNR.
The blue (red) region denotes the $+$ sign ($-$ sign)
with the isovalue of $6.7 \times 10^{-3} |e| $\AA$^{-3}$.
}
\end{figure}

\section*{Results and Discussion}

We have considered the $n$-ZGNR that has $n$ units
of the C-C pair in the $x$-direction per unit cell and repeats along the $y$-axis
with the periodicity of 2.46 \AA. The unit cell of the PVDF also repeats along the
$y$-axis, but with a different equilibrium unit cell length of 2.54 \AA~ which
is 3.15\% longer than that of the ZGNR (Figure 1a).
For convenience, the unit cell size of the PVDF is set to be reduced in the $y$-direction
so that it matches the length of the ZGNR (2.46 \AA) in studying the joint system
(Figure 1b,c).
The unit cell in this system is indicated in Figure 1c.
We confirm that our atomic models with the slightly compressed PVDF on the ZGNR
give essentially the same results as the ones obtained from a fully relaxed geometry
(without compression) in a very large commensurate unit-cell for the PVDF and the ZGNR.
We will discuss this issue later. The vacuum along the $x$-axis is set to be longer than 70 \AA~
in order to avoid the spurious dipole interactions
between the adjacent PVDFs in repeated supercells.
All properties remain the same when the supercell size is doubled, and
the dipole correction is negligible for our purpose.
In determining the most stable position of the PVDF on the ZGNR,
three parameters have to be optimized simultaneously: directions of
the dipole moment of the PVDF, heights of the PVDF from the ZGNR plane, and distances between the PVDF
and the edge of the ZGNR along the $x$-axis.
From a combinatorial search, the structure is found to be stabilized
when the dipole orientation is parallel to the surface of the ZGNR,
the height of the lowest atom of the PVDF from the ZGNR is 2.5 \AA,
and the PVDF lies on the middle position of the ZGNR (Figure 1b,c).

From theoretical~\cite{Duan} and experimental~\cite{Bune,Noda1,Noda2,Choi} studies,
the dipole orientation of the PVDF is known to depend highly on the kind of substrate.
When the substrate is a conductor, the image dipole is induced by the dipole of the PVDF
on conducting substrates and the interaction energy between two dipoles
should be taken into account.
If we regard the PVDF as a single dipole, the energy of perpendicular dipoles is $-2 p^2 / r^3$
from the dipole-dipole interaction energy equation~\cite{Duan},
where $p$ and $r$ denote the magnitude of the dipole moment and
the distance between the dipole of the PVDF and its image dipole, respectively.
On the other hand, the energy of parallel dipoles is $-p^2 / r^3$.
Thus, on the conducting substrate, the perpendicular dipole direction
of the PVDF is energetically favored~\cite{Duan,Choi}.
On the contrary, in the case of an insulating or semiconducting substrate,
for example, KBr~\cite{Noda2} or KCl~\cite{Noda1}, the dipole is aligned parallel
to the surface plane because the image dipole is negligible~\cite{Duan}.
From the calculations, we find that the PVDF favors lying flat to the ZGNR
plane since the ZGNR is semiconducting with its low screening capability.

The distance from the ZGNR to the lowest atom of the PVDF is found to be 2.5 \AA,
and it is 3.5 \AA~ from the ZGNR to the C-C bond of the PVDF (Figure 1b).
From the fully relaxed geometries, we can infer that the PVDF weakly binds
to the ZGNR and the PVDF affects the electronic structure of the ZGNR not
through chemical bonding between the PVDF and the ZGNR, but through
the electronic potential induced by the strong dipole moment.
The structure shown in Figure 1c, where the C-C bonds
of the PVDF and those of the ZGNR cross each other at the middle of the ZGNR, has been
found to a ground state.
Even if the PVDF deviates its ground state geometry slightly, we find that no essential
variation occurs in the electronic structures due to the weak bonding between the PVDF and the ZGNR.
It is also noticeable that, with the PVDF, the antiferromagnetic spin configuration
of the ZGNR is still more stable than the ferromagnetic one.
For a single PVDF on the 8-ZGNR, the interedge antiferromagnetic ordering is
more stable by 2.1 meV per edge atom.
The isosurface of the charge density difference between the up-spin and
the down-spin ($\rho_{\uparrow}(r) - \rho_{\downarrow}(r)$) in this case
is presented in Figure 1d.

Now, we analyze the band structure of this joint system, a single PVDF on the 8-ZGNR.
The pristine 8-ZGNR shows a small band gap of 0.3 eV for both spins.
With one PVDF deposited on the 8-ZGNR, the band structure of the ZGNR is
changed as if it were under external transverse electric fields~\cite{Son2} (Figure 2).
The electric potential energy is raised at the fluorine side and lowered
at the hydrogen side of the PVDF.
Since the dipole field decays slowly in space, the potential drop induced
by this dipole field takes place over a large area which can cover both edges
of the ZGNR.
The energy eigenvalues of the ZGNR are also modified by this potential variation
so that the band gap of the, say, up-spin state is reduced to 0.18 eV,
and that of the down-spin state is increased up to 0.41 eV (Figure 2).
We can estimate the potential profile generated by a single PVDF chain
on the ZGNR by averaging the potential energies in the $y$-axis with the position
of the ZGNR fixed in $z$ direction.
Its drop across two edges of the PVDF is 1.2 V and that of the 8-ZGNR is 0.8 V.
According to previous works~\cite{Son2}, such a band gap change corresponds
to the case of homogeneous electric field of 0.05 V/\AA~ applied to the 8-ZGNR,
that is, the total potential difference between two edges of the 8-ZGNR is 0.89 V
( = 0.05 V/\AA~ $\times$ 17.85 \AA, where 17.85 \AA~ is the length of the 8-ZGNR),
which agrees reasonably well with the present calculation of 0.8 eV.
Thus, the change of the bandgap mainly originates from the potential difference
between two edges of the ZGNR.
When the electric field is applied, the energy eigenvalues of the states
localized on the left side of ZGNRs are pulled up, and those
on the right side are pushed down~\cite{Son2}.
Then, the energy gap of the up-spin state decreases, and that of
the down-spin increases.
We conclude that the energy eigenvalues of ZGNRs are modified
because of interplay between the antiferromagnetic interedge ordering and
the dipole field of the PVDF.

\begin{figure}[t]
\centering
\includegraphics[width=1.00\columnwidth]{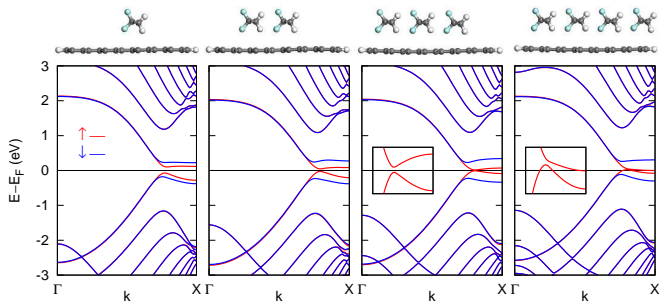}
\caption{
The electronic band structures near the Fermi level.
From left to right, the number of the PVDF chains is increased from one to four.
Due to the potential drop by PVDFs, the band gap of the up-spin decreases,
and that of the down-spin increases. The red (blue) line denotes the band structure
of the up (down) -spin as a function of $k$. (They overlap in most places.)
The inset displays the band structure in the range of $| E - E_{F} |$ $<$ 100 meV
and $0.7\pi \le ka \le \pi$.
In all figures, the Fermi level ($E_F$) is set to zero.
}
\end{figure}

Before proceeding further, we check the lattice mismatch problem
between the PVDF and the ZGNR in the unit cell.
As we mentioned before, the lattice of the PVDF is reduced to that
of the ZGNR in the $y$-axis.
To verify the effect of the compressed PVDF on the ZGNR, same calculations are performed
for a large supercell amounting to 32 unit-cells of the ZGNR
and 31 unit-cells of the PVDF, in which the bonds experience almost no strain.
Very little change of the band structure is observed with a different width of
the PVDF along the periodic direction.
Hence, the role of the PVDF is only to generate the electric field
by its dipole moment, and the position of the PVDF or the reduced cell size does
not appreciably influence the electronic and magnetic properties of the system.
In addition, a typical domain size of the ferroelectric thin film of the PVDF is around 200 nm so that,
if considering a nanoscale channel device geometry, the effect of phase changes along the PVDF
would not be serious in the present case.~\cite{Ducharme}

Next, we increase the number of PVDFs deposited on the 8-ZGNR one by one (Figure 2).
Dipole moments of PVDFs are all aligned in the same direction.
The total dipole moment of PVDFs is increased, and the potential drop
across the ZGNR becomes 1.26, 2.0, and 3.0 V if the number of PVDFs
is 2, 3, and 4, respectively, in the 8-ZGNR case.
(The local potential drop between two edges of PVDFs increases to 2.1, 3.4,
and 5.0 V as increasing the number of PVDFs.)
The band gap of the up-spin state is decreased while that of the down-spin state
increases as the number of PVDFs increase.
Eventually, if four PVDFs are coated on the 8-ZGNR, the system turns to a half-metal.
The up-spin states cross the Fermi level while the down-spin states still have
a band gap of 0.37 eV.
Four PVDFs generate enough of a potential difference (3.0 eV) between two edges
of the ZGNR to give rise to the half-metallicity of the ZGNR.
With low defect concentrations at the edges, the half-metallic nature
of ZGNRs is already shown to be robust under the applied external transverse electric field in the previous work.~\cite{Son2}
Since PVDFs here play the same role as do external fields in ref 11, we note that the induced half-metallicity of
the ZGNRs will occur as long as weak edge defects or imperfections do not alter the magnetic ordering at the edges.

\begin{figure}[t]
\centering
\includegraphics[width=0.70\columnwidth]{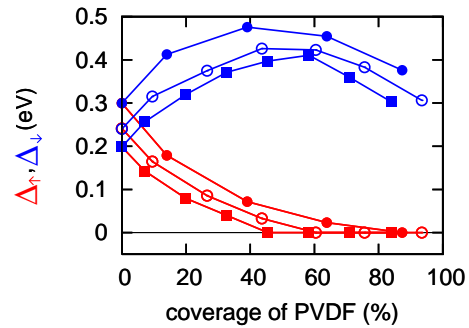}
\caption{
Dependence of half-metallicity on different sizes of the ZGNR and the PVDF.
$\Delta_{\uparrow}$(red) denotes the direct bandgap of the up-spin,
and $\Delta_{\downarrow}$(blue) the gap of the down-spin as a function of the coverage
of PVDFs for the 8-ZGNR (filled circles), 12-ZGNR (open circles), and 16-ZGNR (squres).
When the coverage reaches $16.0 / w \times 100 (\%)$, where $w$ is the width of the ZGNR
in angstrom, the ZGNR becomes half-metallic.
}
\end{figure}

Our calculations indicate that four PVDFs create a potential drop of 3.0 V
across the $n$-ZGNR irrespective of the width ($n$ value) of ZGNRs
(for $n$ $\ge$ 8, tested up to $n$ = 32). The $n$-ZGNR becomes
half-metallic when the potential drop across the system generated by PVDFs
reaches 3.0 V, which has been suggested to be the critical value
for half-metallicity in the previous work~\cite{Son2}.
In terms of the percentage coverage (\textit{i.e.}, the area of covered PVDFs divided by the
area of the ZGNR), the critical value for the transition to half-metal can be calculated by
the width of four PVDFs (16.0 \AA) divided by the width of the ZGNR ($w$),
$16.0 / w \times 100 (\%)$ ($w$ is in angstrom).
This can be regarded as a scaling behavior in the PVDF coverage and the width
of ZGNRs as follows, similar to the previous study on scaling in the strength
of transverse electric fields and the width of the ZGNR~\cite{Son2} (Figure 3).

Finally, another advantage of our proposal compared with others lies
in the fact that the half-metallic state can return to the insulating state
by changing of the dipole orientation of PVDFs.
If the dipole direction of PVDFs is set perpendicular to the ZGNR plane
by a perpendicular external electric field as in the usual experiments
with back gates, then the potential drop across the ZGNR vanishes
and the half-metallicity disappears as well.
We have calculated variations of the total energy
by rotating a single PVDF on ZGNRs.
We have found that, without a perpendicular electric field, the perpendicular
dipole direction of the PVDF to the ZGNR is also at the local energy minimum
(or quasi-stable configuration), 30 meV higher than the ground state
(the parallel dipole direction).
Upon application of perpendicular electric fields, the energy differences
between parallel and perpendicular dipole configurations are decreased
and eventually, with an electric field of $\sim$0.07 V/\AA, the perpendicular
dipole direction becomes the ground state.
An energy barrier for $90^{\circ}$ rotation of the PVDF on the ZGNR, that is,
parallel to perpendicular direction change, is 96 meV per unit cell
without an electric field and decreases to $\sim$77 meV with an electric field
of 0.07 V/\AA.
Therefore, it is practically possible to switch between half-metallic
and insulating states by applying a perpendicular gate field greater
than 0.07 V/\AA.

\section*{Conclusion}

We have proposed a method to construct the half-metallic
ZGNRs by depositing the ferroelectric material
PVDFs on them.
The PVDFs generate the electrostatic potential on the ZGNR because of their strong
dipole moments.
When the number of PVDFs coated on ZGNRs increases beyond a critical value,
the system becomes be half-metallic.
The dipole direction of PVDFs is changeable by an external electric field,
and switching between half-metallic and insulating states in ZGNRs is made possible.

\section*{Theoretical Methods}

We performed the first-principle calculations based on the density-functional
theory (DFT) within the local spin density approximation (LSDA)~\cite{Soler}
and spin polarized general gradient approximation (GGA) with the Perdew-Burke-Ernzerhof (PBE)
functional~\cite{PBE}, respectively, by using the SIESTA package~\cite{Soler}.
%The local density functional parametrized by Ceperley and Alder~\cite{Soler}
%was used for the exchange-correlation potential of the electrons system.
The standard norm-conserving
Troullier-Martins pseudopotentials~\cite{tmpseudo} were employed,
and split-valence double-$\zeta$ plus polarization
basis~\cite{Soler} was used.
We have chosen a 400 Ry energy cutoff for a real space mesh size and 96 $k$-points,
uniformly distributed in the 1D Brillouin zone. All edges were saturated with hydrogen
atoms and relaxed by conjugate gradient minimization until the maximum force was less
than 0.04 eV/\AA.
To overcome an intrinsic error due to the pseudoatomic orbital basis set in the weakly
binding system, the basis set superposition error (BSSE)
was corrected using a counterpoise procedure~\cite{BSSE,Stone}.
In calculations with the PBE functional, we added a pairwise interatomic $C_{6} R^{-6}$ term ($E_{\rm{vdW}}$)
to the PBE-DFT energy in order to include the van der Waals (vdW) interaction~\cite{Grimme, Barone, Scheffler}.
To obtain accurate vdW energies, the effective $C_{6}$ coefficients of each atom in the systems
were obtained by using a recent theoretical method exploiting ground state electron density
from A. Tkatchencko and M. Scheffler~\cite{Scheffler}. From the calculation including PBE and vdW corrections, 
we found that there is no significant difference between atomic and electronic structures 
based on LDA and those based on PBE and vdW corrections. The comprehensive methods and comparisons
are in the Supporting Information. Hence, in the article, all binding energies, relative distances, and orientations between PVDFs and ZGNRs
and band structures in the paper are obtained within the LDA calculations.

\section*{Acknowledgement}              % ACS format
Y.-W.S. thanks B. \"Ozyilmaz for discussions.
Y.-L.L., S.K., C.P. and J.I. were supported by the KOSEF funded
by the Korea Government MEST (Center for Nanotubes and Nanostructured Composites),
and the basic Research Fund No. KRF-2006-341-C000015.
Y.-L.L. acknowledges a Seoul Science Scholarship.
Y.-W.S. was supported by NRF, funded by the MEST (Grant No. R01-2007-000-10654-0
and Quantum Metamaterials Research Center, No. R11-2008-053-01002-0).
Computational resources were provided by KISTI and the KIAS Linux cluster system.

\section*{SUPPLEMENTARY INFORMATION}

It is well known that the long range interaction tail~\cite{Stone} is not included
in local-density approximation (LDA)~\cite{Soler} and generalized gradient approximation
(GGA)~\cite{Perdew,PBE} to exchange-correlation energy functional
in density functional theory (DFT).
Because either LDA or GGA are based on the uniform electron gas system,
they fail to describe weak interactions between electrons in sparse materials
such as layered systems, liquid crystals, polymers, proteins, and biomolecular surfaces~\cite{Ortmann}.
Nevertheless, LDA and GGA calculations provide upper and lower limits of bond lengths,
binding energies, and energy band gaps. They can be useful for reference data sets.
Moreover, many excellent methods~\cite{Wu,Grimme,Zimmerli,Ortmann,Barone,Scheffler,Langreth}
which can describe the long range interaction or van der Waals (vdW) forces
within DFT calculations are proposed.

In our main article, Poly (Vinyliden Fluoride) (PVDF) can be deposited
on the zigzag graphene nanoribbon (ZGNR) with the parallel dipole direction
to the graphene plane.
From binding energies and bond lengths between them from LDA and GGA approximations,
we infer that PVDFs weakly bind to ZGNR,
therefore the vdW correction should be included.
In this supporting information, we present binding energies and bond lengths
from LDA and GGA calculations when one PVDF is deposited on the 8-ZGNR.
The vdW correction is considered by adding a pairwise interatomic $C_{6} R^{-6}$ term ($E_{vdW}$)
to the DFT energy~\cite{Grimme, Barone, Ortmann, Scheffler}.
Finally, comparisons with LDA and GGA for the ground state under the external
electric fields are also discussed.

\begin{figure}[t]
\centering
\includegraphics[width=7cm]{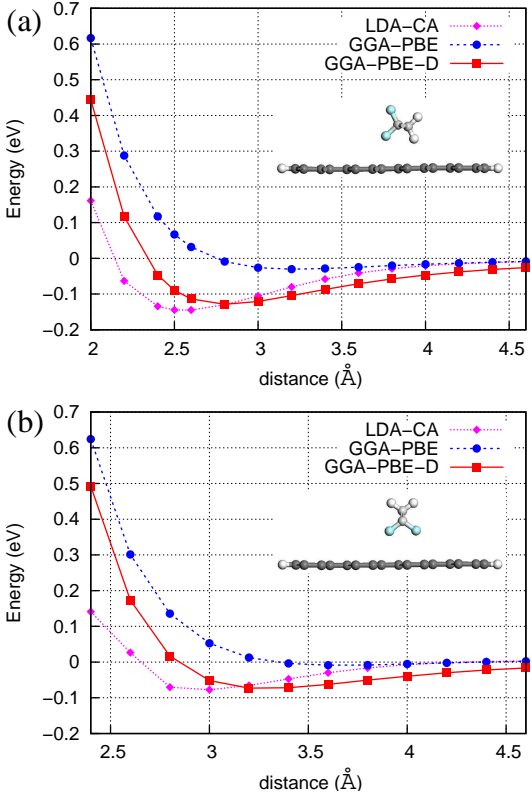}
\caption{
Binding energies as increasing the distance between 8-ZGNR and PVDF from 2.0 \AA~ to 4.6 \AA.
We carried out LDA (dotted line), GGA (dashed line) and vdW corrected GGA,
it is called GGA-D (solid line). (a) With parallel dipole direction
of PVDF to 8-ZGNR and (b) with perpendicular dipole direction of PVDf to 8-ZGNR.
}
\end{figure}

We performed LDA and GGA calculations when depositing one PVDF on 8-ZGNR
as changing the distance between the two from 2.0 \AA ~to 4.6 \AA~
and the dipole direction of PVDF, parallel and perpendicular to 8-ZGNR.
In order to reduce an intrinsic error due to the small number of basis sets,
basis set superposition error (BSSE),
we consider corrections for the error by introducing ghost orbital basis sets~\cite{Stone,BSSE}.
We compare binding energies at each distance with LDA and GGA, respectively.
Binding energies are energy difference between total system (PVDF deposited on 8-ZGNR)
and separated system (8-ZGNR, PVDF).
They are calculated by following formula,
$E_{bind,CP}=E($8-ZGNR$ + $PVDF$)-E($8-ZGNR, ghost atoms of PVDF$)-E($PVDF, ghost atoms of 8-ZGNR$)$,
CP means `counterpoise' correction~\cite{Stone}, and ghost atoms have only pseudo atomic orbitals
without potential~\cite{BSSE}.

In Fig. 1(a), dotted (dashed) line represents binding energies calculated by LDA (GGA)
with parallel dipole of PVDF to 8-ZGNR.
The binding energy of PVDF on 8-ZGNR from LDA (GGA) is 145 meV (30 meV) and
the distance between them is 2.5 \AA ~(3.2 \AA) (parallel dipole direction to the surface,
see inset of Fig. 1(a)).
These values agree well with
tendency of LDA and GGA calculations, i.e., overbinding in LDA and underbinding in GGA.
When dipole direction of PVDF is perpendicular to 8-ZGNR, binding energies are
also plotted in Fig. 1(b) under LDA and GGA calculations.
The binding energy is 77 meV and 8 meV, and the distance between PVDF and 8-ZGNR
is 3.0 \AA~and 3.6 \AA~ for LDA and GGA calculations, respectively.
From those results, we confirm that parallel dipole of PVDF is more stable than
perpendicular dipole of PVDF on 8-ZGNR irrespective of the approximations.

\begin{figure}[t]
\centering
\includegraphics[width=7cm]{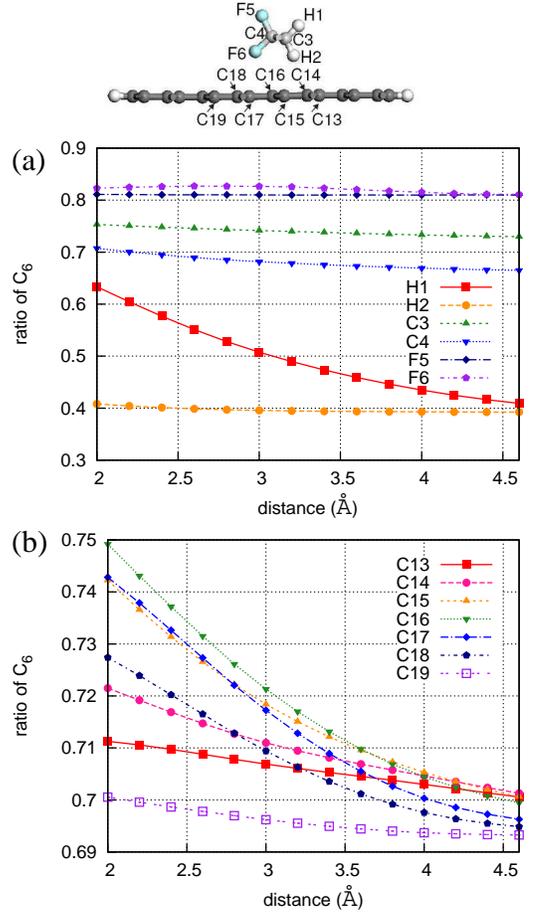}
\caption{
The ratio of $C_6^{eff}$ to $C_6^{free}$ for each atom in PVDF and 8-ZGNR as increasing
distance between them. (a) The ratio
for all atoms in PVDF, and (b) for some atoms in 8-ZGNR which are close to PVDF.
	The number index of atoms in each figure is denoted in the atomic model above the panels.
}
\end{figure}

In order to consider the vdW interaction between PVDF and 8-ZGNR,
we added vdW correction energy $E_{vdW}$ to the DFT total energy ($E_{DFT-D}$),
\begin{equation}
E_{DFT-D}=E_{KS-DFT}+E_{vdW},
\end{equation}
where $E_{KS-DFT}$ is the usual self-consistent Kohn-Sham energy.
The vdW energy ($E_{vdW}$) is given by
\begin{equation}
E_{vdW}=-\frac{1}{2} \sum_{A,B} f_{dmp}(R_{AB}, R_{A}^{0},R_{B}^{0})C_{6AB}R_{AB}^{-6}.
\end{equation}
Here, $R_{AB}$ is the distance
between atoms $A$ and $B$. $C_{6AB}$ is the $C_{6}$ coefficient between atom $A$ and $B$ which can be written
as $C_{6AB}=\frac{2C_{6AA}C_{6BB}}{\frac{\alpha^0_B}{\alpha^0_A}C_{6AA}+\frac{\alpha^0_B}{\alpha^0_A}C_{6BB}}$
where $C_{6AA(BB)}$ and $\alpha^0_{A(B)}$ are homonuclear $C_6$ coefficient and static
polarizability of atom A(B).
$R_{A}^{0}$ and $R_{B}^{0}$ are the vdW radii.
The $R_{AB}^{-6}$ singularity at small distances is eliminated
by the short-ranged damping function $f_{dmp}(R_{AB},R_{A}^{0},R_{B}^{0})$.
\begin{equation}
f_{dmp}(R_{AB},R_{AB}^0)=\frac{1}{1+\exp[-d(\frac{R_{AB}}{s_R R_{AB}^0}-1)]},
\end{equation}
where $R_{AB}^0=R_A^0+R_B^0$. $d$ and $s_R$ are free parameters.
To calculate $E_{vdW}$,
we use a method proposed by A. Tkatchenko and M. Scheffler~\cite{Scheffler}.
In the method, we can calculate the effective parameters, $C_{6}^{eff}$ and $R_{eff}^{0}$ corresponding
to $C_{6AB}$ and $R^0$ respectively from the ground state electronic density. This improves the description of
weakly bonded systems compared with empirical ones~\cite{Scheffler}.

We added this vdW correction to GGA calculations with Perdew-Burke-Ernzerhof (PBE)~\cite{PBE}
functionals using well-known free parameters, $d=20$ and $s_R=0.94$~\cite{Scheffler},
$C_6^{free}$, $R^0_{free}$, and $\alpha_A^0$, shown in Table I.
The summation in Eq. (2) is done for atoms within 500 \AA.

\begin{table}[t]
\centering
\begin{tabular}{|c|c|c|c|}
\hline
 &
$C_{6AA}^{free}$~\cite{Chu} &
$R_{free}^0~\cite{Bondi}$ &
$\alpha_0~\cite{Chu,Yan}$ \\
\hline
H &6.5 & 1.2 & 4.5 \\
C &46.6 &1.70 & 12.0 \\
F &9.5 & 1.47 & 3.8 \\
\hline
\end{tabular}
\caption{
Free-atom $C_6$ coefficients (hartree $\cdot$ bohr$^6$) from Chu~\cite{Chu},
vdW radius $R_{free}$ (\AA) from Bondi~\cite{Bondi} and static polarizability
from Chu~\cite{Chu} and Yan~\cite{Yan}.
}
\end{table}

The binding energy including the BSSE correction~\cite{BSSE},
$E_{bind-D}=E($8-ZGNR$+$PVDF$)_{DFT-D}-E($8-ZGNR$)_{DFT-D}-E($PVDF$)_{DFT-D}$, 
is plotted as solid line in Fig. 1(a) and (b).
It gives the binding energy of 129 meV and the distance of 2.8 \AA~in (a) and 73 meV 3.2 \AA~in (b).
They are between LDA and GGA limits and quite close to the LDA results.
From all results, nevertheless vdW correction changes binding energy
and bond length between PVDF and 8-ZGNR, it is not a crucial part in determining the bonding nature
and electronic structures.

Additionally, the effective $C_6$ coefficients, $C_6^{eff}$ in the system decrease
from those of free atoms and changes when system configurations are altered.
If we change the height between PVDF and ZGNR,
the $C_6^{eff}$ between the nearest pair atoms belong to each material decreases significantly
while those for well separated pair atoms decreases slightly [Fig. 2].
Because different hybridizations states for atoms in each system give various $C_6$ coefficiensts,
all $C_6$ values are changed, respectively.
The ratio of $C_6^{eff}$ to $C_6^{free}$ is plotted in the Fig. 2(a) and (b).
The ratio for all PVDF atoms are shown in Fig. 2(a) and the ratio for some 8-ZGNR atoms
which are close to PVDF are presented in Fig. 2(b).
The ratios are in the range of 0.37 to 0.6 for Hydrogen atoms, 0.64 to 0.75 for Carbon atoms,
and about 0.81 for Fluorine atoms. Especially, the closest atom between PVDF and 8-ZGNR,
H1 (Fig. 2(a)), has large variation of the $C_6$ ratio, and carbon atoms at both edges of 8-ZGNR
have constant ratio of 0.64.

Finally, the vdW correction is also performed under the external electric fields.
We found that the LDA results are still trustful.
Based on electrostatic forces, the dipole of PVDF tends to
align with parallel direction to external electric fields.
With the electric fields of 0.1 V/\AA~ whose direction is perpendicular to ZGNR surface,
the perpendicular PVDF to ZGNR surface is more stable than the parallel one.
The energy differece between perpendicular and parrallel direction of PVDF is 14 meV and 8 meV
in LDA and vdW corrected GGA, respectively.

In conclusion, we carried out LDA, GGA, and vdW corrected GGA calculations for a single
PVDF on 8-ZGNR.
Our vdW corrected GGA-PBE calculatons show that binding energies, distances between PVDF and ZGNR,
and the response under the external electric fields lead to the same conclusion to ones
from LDA calculations with very small quantitative differences.

\end{document}